\documentclass{emulateapj}
\newcommand{\kms}{\ensuremath{\rm{km\,s^{-1}}}}

\newcommand{\ha}{{H}{$\alpha$}}
\newcommand{\hi}{\ion{H}{1}}

\newcommand{\pa}{{\rm PA}}
\renewcommand{\vec}[1]{\mbox{\boldmath $ #1 $}}

\newcommand{\D}{\ensuremath{D}}

\newcommand{\eg}{e.g.}

\newcommand{\chisq}{\ensuremath{\chi_{\rm r}^2}}
\newcommand{\chisqm}{\ensuremath{\chi_{\rm r,min}^2}}
\newcommand{\dn}{\ensuremath{D_n}}

\newcommand{\xn}{\ensuremath{x_n}}
\newcommand{\yn}{\ensuremath{y_n}}
\newcommand{\ptn}{\ensuremath{(\xn,\yn)}}
\newcommand{\xe}{\ensuremath{x_e}}
\newcommand{\ye}{\ensuremath{y_e}}
\newcommand{\pte}{\ensuremath{(\xe,\ye)}}
\newcommand{\rn}{\ensuremath{r_n}}
\newcommand{\sn}{\ensuremath{\sigma_n}}
\newcommand{\wk}{\ensuremath{w_{k,n}}}
\newcommand{\wkb}{\ensuremath{ \{ \wk \}}}
\newcommand{\wj}{\ensuremath{w_{j,n}}}
\newcommand{\vk}{\ensuremath{V_k}}
\newcommand{\vkb}{\ensuremath{ \{ \vk \}}}
\newcommand{\sumk}{\ensuremath{\sum_{k=1}^K \wk \vk}}
\newcommand{\dof}{\ensuremath{\nu}}
\newcommand{\sini}{\ensuremath{\sin{i}}}
\newcommand{\dv}{\ensuremath{\delta V}}
\newcommand{\rout}{\ensuremath{r_{\rm max}}}
\newcommand{\res}{\ensuremath{\Delta V_n}}

\newcommand{\averes}{\ensuremath{\langle|\res|\rangle}}

\newcommand{\eps}{\ensuremath{\epsilon_d}}
\newcommand{\pdsky}{\ensuremath{\phi_d^{\prime}}}
\newcommand{\xc}{\ensuremath{x_c}}
\newcommand{\yc}{\ensuremath{y_c}}
\newcommand{\cen}{\ensuremath{(\xc,\yc)}}
\newcommand{\pbsky}{\ensuremath{\phi_b^{\prime}}}

\newcommand{\Vsys}{\ensuremath{V_{\rm sys}}}
\newcommand{\eISM}{\ensuremath{\Delta_{\rm ISM}}}
\newcommand{\mprim}{\ensuremath{m^{\prime}}}

\newcommand{\pbp}{\ensuremath{\phi_b}}
\newcommand{\td}{\ensuremath{\theta}}
\newcommand{\tb}{\ensuremath{\theta_b}}
\newcommand{\Vmod}{\ensuremath{V_{\rm model}}}
\newcommand{\Vrot}{\ensuremath{\bar V_t}}
\newcommand{\Vrad}{\ensuremath{\bar V_r}}
\newcommand{\Vbit}{\ensuremath{V_{2,t}}}
\newcommand{\Vbir}{\ensuremath{V_{2,r}}}

\newcommand{\Vrotr}{\ensuremath{\Vrot(r)}}
\newcommand{\Vradr}{\ensuremath{\Vrad(r)}}
\newcommand{\Vbitr}{\ensuremath{\Vbit(r)}}
\newcommand{\Vbirr}{\ensuremath{\Vbir(r)}}

\newcommand{\rotrad}{radial}
\newcommand{\rotbi}{bisymmetric}

\newcommand{\dVrotr}{\ensuremath{\Delta \Vrot(r)}}

\newcommand{\gal}{NGC~2976}
\defcitealias{s03}{SBLB}

\long\def\Ignore#1{\relax}

\slugcomment{To appear in ApJ}
\shorttitle{Non-Circular Motions in Disk Galaxies}
\shortauthors{Spekkens \& Sellwood}

\begin{document}

\title{Modeling Non-Circular Motions in Disk Galaxies: \\ Application to \gal}

\author{Kristine Spekkens\altaffilmark{1,2} \&  J. A. Sellwood\altaffilmark{2}}

\altaffiltext{1}{National Radio Astronomy Observatory (NRAO). NRAO is a facility of the National Science Foundation operated under cooperative agreement by Associated Universities, Inc.}
\altaffiltext{2}{Department of Physics and Astronomy, Rutgers, the State University of New Jersey, 136 Frelinghuysen Road, Piscataway, NJ, 08854.}

\email{spekkens, sellwood@physics.rutgers.edu}

\begin{abstract}
We present a new procedure to fit non-axisymmetric flow patterns to
2-D velocity maps of spiral galaxies.  We concentrate on flows caused by
bar-like or oval distortions to the total potential that may arise
either from a non-axially symmetric halo or a bar in the luminous
disk.  We apply our method to high-quality CO and \ha\ data for the nearby, low-mass spiral \gal\ previously obtained by Simon et al., and find that a
bar-like model fits the data at least as well as their model with
large radial flows.  We find supporting evidence for the existence of
a bar in the baryonic disk.  Our model suggests that the azimuthally averaged central
attraction in the inner part of this galaxy is larger than estimated
by these authors.  It is likely that the disk is also more massive,
which will limit the increase to the allowed dark halo density.
Allowance for bar-like distortions in other galaxies may either
increase or decrease the estimated central attraction.
\end{abstract}


\section{Introduction}
\label{intro}
One of the first steps toward understanding the
formation and evolution of galaxies is a determination of the 
radial distribution
of mass within a representative sample of systems.  The rotational
balance of stars and gas in the plane of a disk galaxy offers a
powerful probe of its mass distribution, and has
been widely exploited \citep{sofue01}.

When the motions of these tracers are consistent with small departures
from circular orbits, the determination of the rotation curve (more precisely, 
the circular orbital speed profile) is
straightforward.  However, it has long been known \citep[\eg][]{bosma78}
that large non-circular motions driven by bar-like or oval
distortions, warps, or lopsidedness are common features in galaxy
velocity maps, which complicate the determination of the radial mass
profile.  Yet the observed flow pattern contains a great deal of
information about the mass distribution, which we wish to extract from
the data.

Since galaxies with closely flat, nearly axisymmetric disks are the
exception, it is desirable to be able 
to estimate a mass profile in
the more general cases.  A number of techniques, which we review in
\S\ref{theory}, already exist for this purpose.  A procedure for
dealing with a warped disk has been successfully developed \citep[\eg][]{begeman87} from the first simple tilted ring analyses 
\citep[\eg][]{rogstad74}, and is now widely used.

Non-axisymmetric distortions to the planar flow can always be
described by an harmonic analysis.  But the approach pioneered by
\citet{franx94} for interpreting the resulting coefficients 
embodies epicycle theory, which is valid only for
small departures from circular orbits and may give misleading results
if the observed non-circular motions are not small compared with the
circular orbital speed.  A number of authors (see \S\ref{theory} for
references) appear to find significant radial flows with speeds that
rival the inferred mean orbital motion.  Such flows violate the 
assumption of small
departures from circular motion, are physically not well motivated,
and the results are hard to interpret.

We therefore propose here a new technique for fitting a general
non-axisymmetric model to the velocity field of a galaxy that allows for
large non-circular motions.  We develop and apply the method
specifically for the case of bar-like or oval distortions, but the
procedure is readily generalized for potentials having other azimuthal
periodicities.

Our simple kinematic model, which we describe in \S\ref{technique},
yields both the mean orbital speed and the amplitudes of the
non-circular streaming velocities.  It is successful because (1) we
invoke a straight bar-like distortion to the potential, (2) we do not
need to assume small departures from circular motion, and (3) we fit
to the entire velocity field at once.

We apply our method (\S\ref{n2976}) to the high-quality velocity maps
of \gal\ that were previously presented by \citet[][hereafter SBLB]{s03}, 
and find that it suggests a significantly different radial mass
profile from that deduced by those authors.  We show (\S\ref{discuss})
the reason for this difference, and argue that a bisymmetric
distortion is both a more reasonable physical model, and that it is
supported by independent evidence of a bar in this galaxy.

\section{Modeling Non-Axisymmetric Flows}
\label{theory}

\subsection{Mathematical preliminaries}
\label{math}
The velocity of a star or an element of gas in the plane 
of the disk of a galaxy
generally has two components at each point: tangential, $V_t$, and
radial, $V_r$, relative to any arbitrary center, most conveniently the
kinematic center.  Without approximation, each component can be
expressed as a Fourier series around a circle of radius $r$ in the
disk plane:
\begin{equation}
V_t(r,\theta) = \Vrotr + \sum_{m=1}^\infty V_{m,t}(r)\cos \left[
m\theta + \theta_{m,t}(r) \right]
\end{equation}
and
\begin{equation}
V_r(r,\theta) = \Vradr + \sum_{m=1}^\infty V_{m,r}(r)\cos \left[
m\theta + \theta_{m,r}(r) \right],
\end{equation}
where the coefficients, $V_{m,t}$ and $V_{m,r}$, and phases relative
to some convenient axis, $\theta_{m,t}$ and $\theta_{m,r}$, are all
functions of $r$.  The quantity $\Vrotr$ is the mean streaming speed
of the stars or gas about the center; 
throughout, we refer to this quantity as the 
mean orbital speed.  The axisymmetric term of the radial
motion, $\Vradr$, represents a mean inflow or outflow in the
disk plane, which gives rise to a ``continuity problem'' if it is large \citep{s05}.

Galaxies are observed in projection, with inclination $i$, about a
major axis, which we choose to define as $\theta=0$ in the above
expansions.  The line-of-sight velocity is the sum of the projected
azimuthal and radial velocities: $V_{\rm obs} = \Vsys + \sin i
(V_t\cos\theta + V_r\sin\theta)$, where $\Vsys$ is the systemic
velocity of the galaxy.  In terms of our Fourier series,
\begin{eqnarray}
\nonumber
V_{\rm obs} & = & \Vsys \\
\nonumber
            &    &+ \sin i\left\{ \Vrot\cos\theta + \sum_{m=1}^\infty
V_{m,t} \cos\theta \cos\left[ m\theta + \theta_{m,t} \right] \right. \\
&  &\left. + \; \Vrad\sin\theta + \sum_{m=1}^\infty
V_{m,r}\sin\theta \cos\left[ m\theta + \theta_{m,r} \right] \right\}.
\label{defVobs}
\end{eqnarray}
Using standard trigonometric relations, this expression can be
rewritten as
\begin{eqnarray}
\nonumber
{V_{\rm obs} - \Vsys \over \sin i} 
& = &  \Vrot\cos\theta \\
\nonumber
& & + \sum_{m=1}^\infty {V_{m,t} \over 2}
\left\{ \cos\left[ (m+1)\theta + \theta_{m,t} \right] \right. \\
\nonumber
& & + \left. \cos\left[
(m-1)\theta + \theta_{m,t} \right]\right\} + \; \Vrad\sin\theta \\
\nonumber
& &+ \sum_{m=1}^\infty {V_{m,r} \over 2} \left\{
\sin\left[ (m+1)\theta + \theta_{m,r} \right] \right. \\
& & - \left. \sin\left[ (m-1)\theta + \theta_{m,r} \right] \right\}.
\label{defVobs2}
\end{eqnarray}
As is well known \citep[\eg][]{canzian93,schoen97,ca97,frid01}, projection
therefore causes velocity distortions with intrinsic sectoral harmonic
$m$ to give rise to azimuthal variations of orders
 $\mprim=m\pm1$ in the corresponding line-of-sight velocities.  Thus
intrinsic distortions at two different sectoral harmonics give rise to
projected velocity features of the same angular periodicity in the
data, complicating the determination of all coefficients in the
expansion.

\subsection{Previous approaches}
\label{previous}
The principal scientific objective of most spectroscopic observations 
of disk galaxies is to extract
the function $\Vrotr$, which should be a good approximation to the
circular orbital speed if all other coefficients on the right-hand side of 
eq.~\ref{defVobs} are small.
With a single slit spectrum along the major axis of the galaxy, one
generally sets $V_{\rm obs} = \Vsys + \Vrotr\sin i$, implicitly
assuming all other terms to be negligible.  In this case, the
inclination must be determined from other data (\eg\ photometry).
Differences larger than measurement errors between the approaching and
receding sides flag the existence of non-circular motions, but
measurements along a single axis do not yield enough information to
determine any other coefficient.  This and other uncertainties
inherent in such deductions are well-rehearsed 
\citep{vbd01,dbb03,swaterscuspcore,rhee04,spekkens05,hayashi06}.
A two-dimensional velocity map, on the other hand, provides much more
information.

Software packages, such as {\it rotcur} \citep{begeman87}, allow one to
fit the velocity field with a single velocity function $\Vrotr$ in
a set of annuli whose centers, position angles (PAs) and inclinations are
allowed, if desired, to vary with radius.  This package is ideal for
the purpose for which it was designed: to determine the mean orbital
speed even when the plane of the disk may be warped.  It works well when
non-circular motions are small, but yields spurious variations of the
parameters when the underlying flow contains non-axisymmetric,
especially bisymmetric, distortions.

\citet{bs03} adopted a different approach. They assumed the plane of the disk to
 be flat, and determined the rotation center, inclination, and PA by fitting a
 non-parametric circular flow pattern to the entire velocity map. Their method 
 averages over velocity distortions caused by spiral arms, for example, but again
 may yield spurious projection angles and mean orbital speeds if there is a
 bar-like or oval distortion to the velocity field over a wide radial range. 
 The {\it rotcurshape} program, recently added to the NEMO \citep{teuben95} package,
 suffers from the same drawback because it also assumes a flat, axisymmetric disk.
 Furthermore, it fits multiple parametric components to a velocity field and thus has
 less flexibility than the \citet{bs03} technique.
 
\citet{franx94} and \citet{schoen97} pioneered efforts to measure and 
interpret the non-axisymmetric coefficients that describe an observed 
velocity field, and expansions up to order $\mprim\sim 3$ are now
routinely carried out \citep[\eg][]{wong04,chemin06,s05,gentile06}. 
  Their
approach assumes departures from circular motion to be small so that
the radial and tangential perturbations for any sectoral harmonic can
be related through epicycle theory.  The technique is therefore
appropriate only when all fitted coefficients are small and the mean
orbital speed is close to the circular orbital speed that balances the 
azimuthally averaged central attraction. \citet{wong04} 
present an extensive discussion of this technique and conclude
that it is difficult to work backwards from the
derived Fourier coefficients to distinguish
between different physical models.

\citet{swaters03}, \citetalias{s03} and \citet{gentile06} report velocity
fields for nearby galaxies that show non-circular motions whose 
amplitude rivals the mean orbital speed at small $r$. Swaters et al. (2003b) note that their \Vrotr\ model fails to reproduce the inner disk kinematics of their target. They correct \Vrotr\ for an isotropic velocity dispersion of 8 km/s, but do not attempt to model the isovelocity twists in their H$\alpha$ velocity field.
%\citet{swaters03}
%simply comment that their \Vrotr\ model fails to reproduce the inner
%disk kinematics of their target.  
\citetalias{s03}
fit the simplest acceptable model to their data: an axisymmetric flow
with just two non-zero coefficients $\Vrotr$ and $\Vradr$. They favor this model over a bar-like distortion partly because the galaxy is not
obviously barred, and partly because they find that the $\mprim =3$
components are scarcely larger than the noise (see also \S\ref{n2976}). The addition of the
radial velocity term \Vrad\ allows a more complicated flow pattern to be
fitted with an axisymmetric model, which significantly improves the
fit to their data.  \citet{gentile06} do detect a radial $\mprim = 3$
component in addition to a strong radial $\mprim = 1$
term in the kinematics that they report, which they conclude ``are
consistent with an inner bar of several hundreds of pc and accretion of 
material in the outer regions''.   Despite finding large non-circular 
motions, the authors
of all three studies nonetheless adopted their derived mean
orbital speed as the ``rotation curve'' of the galaxy, which they
assume results from centrifugal balance with the azimuthally averaged 
mass distribution.

 These deductions are suspect, however.  As we show below
(\S\ref{discuss}.1), a bisymmetric distortion to the flow pattern may
not give rise to a large $\mprim=3$ term in the velocity field, and the
smallness of these terms does not establish the
absence of a strong bisymmetric distortion.  Further, associating the
$\Vrot$ term with the rotation curve is valid only if all departures
from circular motion are small, yet they had found non-circular
velocity components almost as large as the mean orbital speed over a
significant radial range.

Early work on modeling gas flows in barred galaxies is reviewed in Sellwood \& Wilkinson (1993; see their section 6.7).
\citet{weiner01}, \citet{kranz03}, and
\citet{perez04} attempt to build a self-consistent fluid-dynamical model 
of the non-axisymmetric flow pattern.  They
estimate the non-axisymmetric part of the mass distribution from
photometry and try to match the observed flow to hydrodynamic
simulations to determine the amplitude of the non-axisymmetric
components of the potential.  The objective of this, altogether more
ambitious, approach is to determine the separate contributions of the
luminous and dark matter to the potential.  Here, our objective
is more modest: to estimate the mean orbital speed from a velocity map
that may possibly be strongly non-axisymmetric.  Thus their attempt to
separate the baryonic from dark matter contributions seems needlessly
laborious for our more limited purpose.

\section{A New Approach}
\label{technique}

\begin{figure}
\epsscale{1.1}
%\plotone{f1.eps}
\plotone{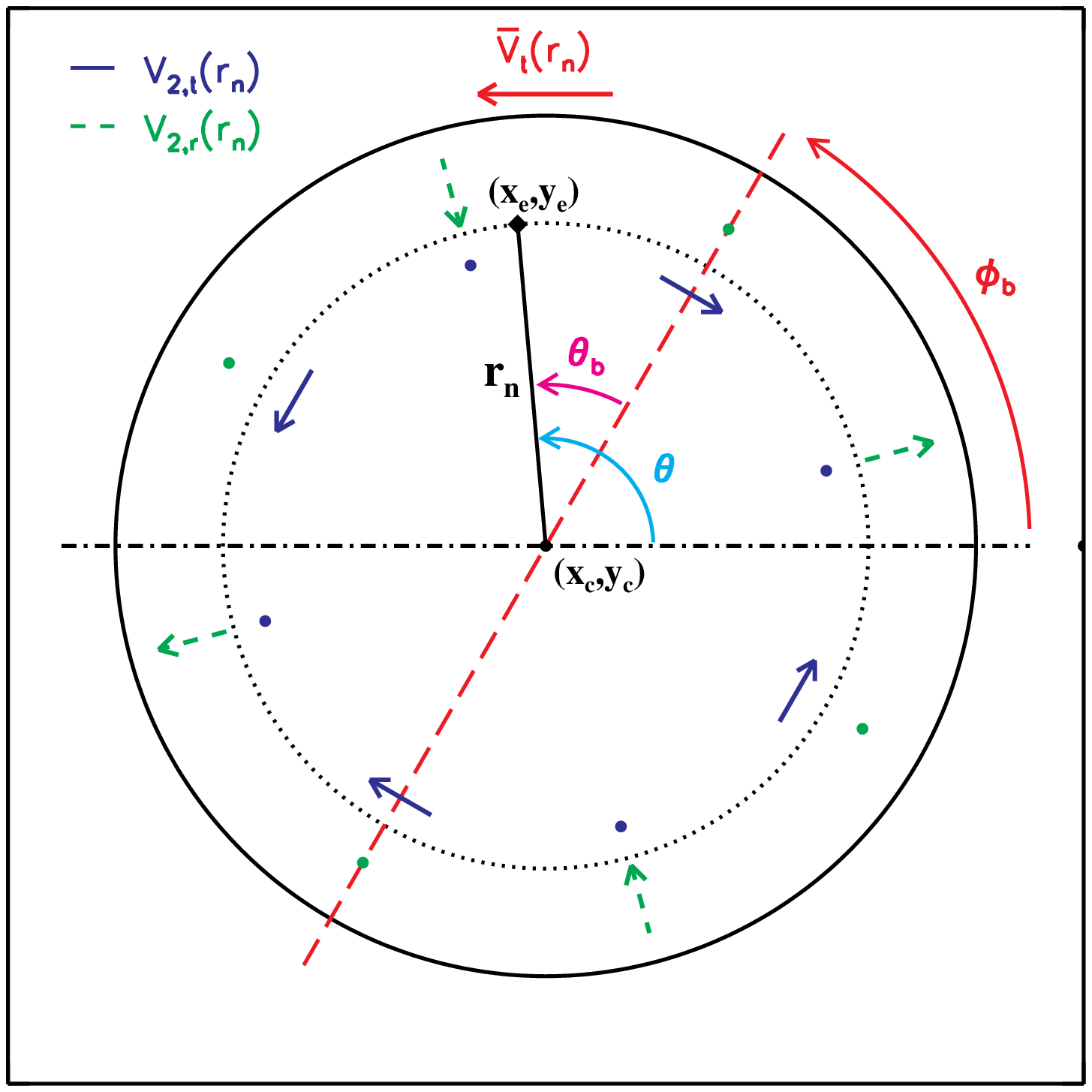}
\caption{Parameter definitions and flow pattern in the disk plane for
the \rotbi\ model (eq.~\ref{bieq}).  The solid circle represents the
largest $r$ included in the model, and the horizontal dash-dotted line is the
major axis of the disk defined by the sky plane.  The long-dashed line 
is the major axis of the bisymmetric distortion, at angle \pbp\ from
the major axis.  The diamond denotes the location \pte\ of a datapoint \dn,
a distance \rn\ from the kinematic center \cen\ and at \pa s \tb\ 
from the bisymmetric
distortion axis and \td\ from the major axis.  The dotted circle shows
the circular orbit of radius \rn\ in the disk, and the axisymmetric
model component $\Vrot(\rn)$ is in the counter-clockwise direction.
The extrema of components $\Vbit(\rn)$ and $\Vbir(\rn)$ are indicated
by solid and dashed arrows, respectively, and large dots at the same distance from \cen\ as each set of arrows denote \pa s where the corresponding component passes through zero.}
\label{setup}
\end{figure}

\subsection{A bar-like distortion}
\label{bisymm}
Our objective is to model non-circular motions in a 2-D velocity map.
Since we do not wish to assume that non-circular motions are small, we
refrain from adopting the epicycle approximation.  However, we do make
the following assumptions:

\begin{itemize}
\item The non-circular motions in the flow stem from a bar-like or
oval distortion to an axisymmetric potential.  We suppose these
motions to be caused by either a non-axially symmetric halo 
in the dark matter or
by a bar in the mass distribution of the baryons.
\item A strong bisymmetric
distortion to the potential, even one that is exactly described by a
$\cos(2\theta)$ angular dependence, can give rise to more complicated
motions of the stars and gas \citep[\eg][]{sw93}.  In particular, the flow may contain
higher even harmonics.  Nevertheless, the $m=2$ terms will always be
the largest, and we therefore begin by neglecting higher harmonics.
\item We assume the bar-like distortion drives non-circular motions
about a fixed axis in the disk plane.  
In a steady bar-like flow, the perturbed parts of
the azimuthal and radial velocities must be exactly out of phase with
each other.  That is, the azimuthal streaming speed is smallest on
the bar major axis and greatest on its minor axis, while radial motions
are zero in these directions and peak, with alternating signs, at
intermediate angles \citep{sw93}.
\item We must assume the disk to be flat, because we require the
predicted $V_{\rm obs}$ from eq.~\ref{defVobs} to have the same
inclination at all $r$.  This assumption is appropriate for spiral
galaxy velocity fields measured within the optical radius, where warps are
rare \citep[\eg][]{briggs90}, and is therefore well-suited to
interpreting kinematics derived from \ha, CO or stellar spectroscopy.
The technique presented here should therefore not be applied to the
outer parts of \hi\ velocity fields, which typically extend well 
into the warp region \citep[\eg][]{broeils97}.
\end{itemize}

A model based on these assumptions predicts the observed velocity at
some general point in the map to be given by eq.~\ref{defVobs}, with
the $m=2$ terms as the only non-axisymmetric terms:
\begin{eqnarray}
\nonumber
\Vmod & = & \Vsys\ + \sini \,\,\left[\, \Vrot\cos{\td}
 - \; \Vbit\cos(2 \tb) \, \cos{\td} \right. \\
 & & \left . - \,\Vbir\sin( 2 \tb) \sin{\td} \,\right] \; .
\label{bieq}
\end{eqnarray}
The geometry in the disk plane is sketched in Fig.~\ref{setup}.  As
above, \td\ is the angle in the disk plane relative to the projected
major axis, which is marked by the horizontal dash-dotted line.  The
major axis of the bisymmetric distortion (or bar for short) lies at
angle \pbp\ to the projected major axis; thus angles relative to the
bar axis are $\tb = \td - \pbp$. We have chosen the phases of the
\Vbit\ and \Vbir\ terms such that both are negative at $\tb=0$ and
they vary with angle to the bar as the cosine and sine of $2\tb$
respectively. Comparing eqs.~\ref{defVobs} and \ref{bieq}, we see that
 $\theta_{2,t} = \pi - 2\pbp$ and $\theta_{2,r}=\pi/2 - 2\pbp$.  

The amplitudes of the tangential and radial components of the
non-circular flow, $\Vbitr$ and $\Vbirr$ respectively, are both
functions of radius.  While it is possible to relate the separate
amplitudes for a given potential, we do not attempt to model the mass
distribution that creates the flow and therefore allow them both to
vary independently.

Circles in the disk plane project to ellipses on the sky, with
ellipticity \eps\ given by $1 - \eps = \cos i$, with a common
kinematic center, \cen.  We use primes to denote projected angles onto
the sky plane.  Thus the projected \pa\footnote{All \pa s
are measured North $\rightarrow$ East.} of the disk major axis
is \pdsky, while \pbsky\ is the \pa\ of the bar major axis in the sky
plane.  These angles are related by
\begin{equation}
\pbsky = \pdsky + {\rm arctan}(\tan \pbp \cos i) \;.
\label{barmajeq}
\end{equation}
In addition to the three velocity functions \Vrotr, \Vbitr\ and \Vbirr, 
the model is therefore
described by the parameters $(\xc,\, \yc,\, \Vsys,\, \eps,\, \pdsky,\,
\pbsky)$. We refer to the model described by eq.~\ref{bieq} as the
\rotbi\ model.

\subsection{Other possible models}
\label{other}
Other models for the flow pattern could readily be derived from
eq.~\ref{defVobs}.

In particular, and solely to facilitate comparison with other work, we also fit a purely axisymmetric model with the coefficents of all
$m>0$ terms set to zero, but retain the $\Vrad$ term.  There is no
undetermined phase angle for this intrinsically axisymmetric model and
the predicted velocity is simply
\begin{equation}
\Vmod = \Vsys + \sini \; \left[ \, \Vrot \cos{\td}
                         + \Vrad\sin{\td} \, \right] \;.
\label{radeq}
\end{equation}
The coefficient $\Vrad$ corresponds to pure radial inflow or
outflow.\footnote{It is not possible to distinguish between inflow and
outflow in this model unless the side of the disk along the minor axis
that is nearest to the observer can be determined independently.}  We
will refer to this as the \rotrad\ model.

Other, more complicated, models could also be fitted to data by
retaining more terms as required, provided that an assumption is
made about the radial dependence of the phases of the non-axisymmetric
perturbations.  The extension of these formulae to include other
velocity field harmonics is straightforward, and we have tried doing
so in some of our analyses (see \S\ref{n2976}).

\subsection{Discussion}
\label{tech_discuss}
If the non-circular motions measured in some spirals 
do stem from bar-like or oval distortions,
then the \rotbi\ model has several advantages over both the \rotrad\
model and also over epicyclic approaches for characterizing these
asymmetries.  Since $\mprim=1$ velocity field components can arise
from either radial flows or a bisymmetric perturbation to the
potential (eq.~\ref{defVobs2}), 
both the \rotbi\ and \rotrad\ models could produce
tolerable fits to the same data.  However, the \rotbi\ model offers a
more direct, unambiguous approach for identifying $m=2$ distortions
than does the \rotrad\ model.

Moreover, interpretations of velocity field harmonics that rely on
epicycle theory \citep{franx94,schoen97,ca97} are applicable only in
the limit of a weak perturbation to the potential, whereas the
components of our \rotbi\ model are not restricted to mild
distortions.  We also note that since the \rotbi\ model imposes a
fixed \pbsky\ on the non-circular flow pattern, it is not sensitive to
$m=2$ perturbations to the potential that are not in phase (such as
spiral patterns).

Finally, the \rotbi\ technique is much simpler than fluid-dynamical
modeling of the velocity field (see \S\ref{previous}), since it does not
require (or yield) a model for the mass distribution.

\subsection{Fitting technique}
\label{minimization}
We attempt to fit the above kinematic models to observational data by
an extension of the minimization procedure devised by \citet{bs03}.
In general, we need to determine the systemic velocity \Vsys,
kinematic center \cen, ellipticity \eps, and
disk \pa\ \pdsky, as well as $M$ unknown radial functions $V_{m,t}$ and $V_{m,r}$
($m=0$ and $m>0$ if desired) and the (fixed) \pa (s),
$\theta_m$, of any non-axisymmetric distortions to the flow.

We tabulate each of the $M$ independent velocity profiles at a set of
concentric circular rings in the disk plane that project to ellipses on the sky
with a common center, \cen.  Once these tabulated values are
determined, we can construct a predicted \Vmod\ at any general point
by interpolation.  We difference our model from the data, which
consist of $N$ line-of sight velocity measurements $\{\dn\}$ with
uncertainties $\{\sn\}$, and adjust the model parameters to determine
the minimum \chisqm\ of the standard goodness-of-fit function \chisq\
with \dof\ degrees of freedom:
\begin{equation}
\chisq=\frac{1}{\nu}\sum_{n=1}^N \left( \frac{\dn - \sumk}{\sn
} \right)^2 \; .
\label{chieq}
\end{equation}
Here, the $K$ elements of $\{\vk\}$ are the values of the tabulated
velocity profiles in the model and the weights, $w_{k,n}$ describe the
interpolation from the tabulated $\vk$ to $\Vmod$ (eq.~\ref{bieq} or
\ref{radeq}) at the position of the observed value \dn.

When $\chisq=\chisqm$, the partial gradient of $\chisq$ with respect
to each $V_j$, where $j$ labels each of the \vk\ in turn, must satisfy
\begin{equation}
{\partial \chisq \over \partial V_j} = -\frac{2}{\nu}\sum_{n=1}^N
\left( \frac{\dn - \sumk}{\sn } \right) \frac{w_{j,n}}{\sn} = 0 \;.
\end{equation}
Rearranging, we find
\begin{equation}
\sum_{k=1}^K \left( \sum_{n=1}^N \frac{\wk}{\sn }\,\frac{\wj}{\sn }
\right) \vk = \sum_{n=1}^N \frac{\wj}{\sn ^2}{\dn } \;,
\label{minchieq}
\end{equation}
resulting in a linear system of $K$ equations for the $K$ unknowns, $\{\vk\}$.  

For a given set of attributes $(\xc,\, \yc,\, \Vsys,\, \eps,\,
\pdsky,\, \theta_m )$ of the projected disk, we compute $\{\vk\}$ by
solving the linear system of eq.~\ref{minchieq}, and use the
resulting $\{\vk\}$ values in eq.~\ref{chieq} to evaluate \chisq.
The best fitting model is found by minimizing
eq.~\ref{chieq} over the parameters mentioned above, which
necessitates recomputing $\{\vk\}$ via eq.~\ref{minchieq} at each
iteration.  Any convenient method may be used to search for the
minimum; we use Powell's direction set method \citep{p92}.

\citet{bs03} use this minimization strategy to extract only the mean
orbital speed $\Vrotr$ from \ha\ velocity fields of spirals in the
\citet{pw00} sample.  In our more general case, the
$M>1$ model profiles are defined by distinct sets of $K'_M$ rings in
the disk plane, and \vkb\ contains all of the velocities from these
profiles: $K = \sum_{M} K'_M$. In other words, adding a velocity
profile (defined in $K'$ rings) to a model increases the rank of the 
matrix in eq.~\ref{minchieq} by $K'$.  The \rotrad\
model has $M=2$, while $M=3$ for the \rotbi\ model.  Further
discussion of \vkb\ and derivations of \wkb\ are given in the
Appendix.

\section{Velocity Field Models of \gal}
\label{n2976}

\begin{figure*}
\epsscale{0.9}
\plotone{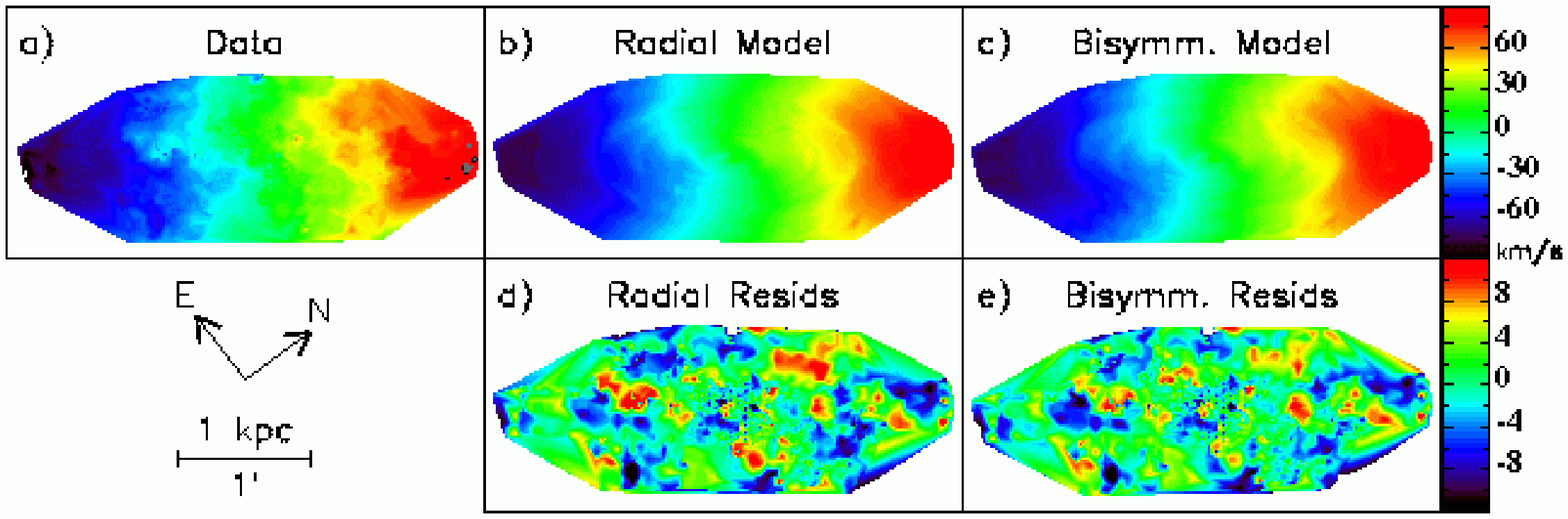}
\caption{Kinematic models of the \gal\ velocity field.  The panels in
the top row show ({\it a}) the observed velocity field $\{\dn\}$ from
\citetalias{s03}, ({\it b}) the optimal \rotrad\ and ($c$) the optimal
\rotbi\ models.  The velocity fields are plotted on the same
colorscale, shown in \kms\ to the right of the top row.  The panels in the
bottom row show residual maps $\{\res\} = \{\dn\ - \sumk\}$ for ($d$)
the \rotrad\ and ($e$) the \rotbi\ models, with the colorscale in
\kms\ for both shown to the right of that row.  The model
velocity fields and residuals have been rotated by $-(\pdsky\ + \pi/2)$
about \cen\ from Table~\ref{fits}.  The data in \ref{models}{\it a} 
are rotated by
the photometric value $-(-37\degr +\pi/2)$ (intermediate to the two
model values) about the photometric center $09^\mathrm{h}\,
47^\mathrm{m}\, 15\fs\, 3$, $67\degr\, 55\arcmin\,00\farcs\,4$
\citepalias{s03}.  The orientation of $\{\dn\}$ (roughly correct for
the models as well) and the map scale are at the bottom left.  The
unusual shape of contoured regions in the maps reflects the locations
of the individual pointings used to construct the \ha\ velocity field
of \gal\ \citepalias{s03}.}
\label{models}
\end{figure*}

To illustrate the technique, we fit our \rotbi\ model to the observed
high-quality velocity field of \gal\ reported by \citetalias{s03}. \gal\ is a nearby, low-mass Sc galaxy with $i \sim
60\degr$. We adopt a distance $\D=3.56\;$Mpc, estimated from the tip
of the red giant branch \citep{kar02}, and convert angular scales to
linear scales using $1\arcsec = 17.3\;$pc.

\citetalias{s03} present \ha\ and CO velocity fields of \gal, with a
spatial resolution\footnote{Throughout, we recompute the linear scales
presented by \citetalias{s03} for consistency with our choice of $D$.}
of $\sim5\arcsec$ ($86\,$pc) and spectral resolutions of $13\,\kms$
and $2\,\kms$, respectively.  They find that the velocity field is not
well-modeled by disk rotation alone.

They report a detailed analysis of these kinematic data in
which the projection geometry of their model rings is determined from
optical and near-IR photometry. They conclude that there is no strong
evidence for a bisymmetric distortion in this galaxy, since all
$\mprim >1$ components of the velocity field are consistent with
noise.  They find that a combination of rotation and pure radial flows
provides an adequate fit.  The amplitude of the inferred radial
velocity profile rivals that of the rotational component for $r
\lesssim 500\,$pc: \gal\ thus exhibits some of the largest
non-circular motions ever detected in a low-mass,
rotationally-supported system.  In their later paper, Simon et al.\
(2005) noted that finding large values of the \Vrad\ term is a
strong indication that a model with an axisymmetric radial flow is
incorrect. They suggest that the non-circular motions in \gal\ stem 
from a triaxial
halo, but their use of epicycle theory relations \citep{schoen97} is inappropriate because 
\Vradr\ is not always small (see their fig. 9). 
We also suspect that a bisymmetric distortion is
responsible for the observed departures from a circular flow pattern.

The \ha\ and CO velocity fields of \gal\ presented in fig.~4 of
\citetalias{s03} were kindly made available to us by J. D. Simon.  Following
these authors, we analyse the kinematics of the two tracers together,
since the data agree within their uncertainties.

We fit the combined velocity field with our \rotbi\ model to examine
whether the departures from circular motion detected by
\citetalias{s03} stem from an $m=2$ distortion to the potential.  In
order to demonstrate that our new technique (\S\ref{technique}) yields
a similar kinematic model to the one obtained by
\citetalias{s03}, we apply our \rotrad\ model to the same dataset, and
compare values of the parameters we obtain as a consistency check on
our method. For completeness, we also attempt to fit the data with
a suite of other models including $m=0$, $m=1$ and $m=2$ distortions 
(see below).

The observations presented by \citetalias{s03} sample the velocity
field of \gal\ out to $r \sim 130\arcsec$ ($2.2\,$kpc) from its
photometric center.  We evaluate the velocity profiles in a maximum of
$K' = 26$ rings, separated by 4\arcsec\ for $r<95\arcsec$ and by up to
10\arcsec\ farther out.  Neither the \rotbi\ model nor the \rotrad\
model yielded reliable constraints on the non-circular components of
the velocity field for $r>100\arcsec$.  We therefore conclude that the
outer part of \gal\ is adequately described by a simple circular flow, and fix
the amplitudes of all coefficients but \Vrot\ to zero beyond that
radius.  This reduces the rank of the matrix (eq.~\ref{minchieq}) by a
few.

To check the validity of our planar disk assumption at the largest
radii probed by the measurements, we compare the disk parameters
derived from fits including different numbers of outer rings.
Specifically, each minimization uses the same ring radii, except that
the outermost ring included is varied in the range $80\arcsec < \rout\
< 135\arcsec$ and velocity measurements at radii beyond $\rout$ are
ignored in the fit.  Models with $\rout \lesssim 112\arcsec$ return
identical disk parameters within the uncertainties, but the optimal
values of \xc, \yc\ and \Vsys\ change substantially when rings at
larger $r$ are added.  We therefore restrict our fits to include only
\dn\ with $\rn < 112\arcsec$ in our final models, as the disk may be
warped farther out.\footnote{The disk geometry and kinematics of \gal\
at $r \gtrsim 1.5\,$kpc will be explored in detail using extant
aperture synthesis \hi\ maps of the system.}

We make an allowance for ISM turbulence by redefining $\{\sn\}$ to be
the sum in quadrature of the uncertainties in the emission line
centroids and a contribution $\eISM=5\,\kms$.  We find that choosing
values of \eISM\ in the range $3\,\kms \lesssim \eISM \lesssim
7\,\kms$ and varying the ring locations and sizes by 2--4\arcsec\ have
little impact on our results.

 In addition to the \rotbi\ and \rotrad\ models of \gal, 
we also fitted models including a lopsided ($m=1$) 
distortion.
The optimal $m=1$ model (including velocity profiles \Vrotr, $V_{1,t}(r)$ and $V_{1,r}(r)$; see eq.~\ref{defVobs}) produced a much less satisfactory fit to the 
data than either the \rotbi\ or the \rotrad\ model. Adding a radial flow
term \Vradr\ yielded optimal parameters identical to those of 
the \rotrad\ model, with the $m=1$ components consistent with zero. 
The insignificance of a lopsided component, which we
conclude from these fits, is consistent with our result below that
the kinematic and photometric centers of \gal\ are coincident within the
errors (see also \citetalias{s03}).  We also attempted to fit
$m=0$ and $m=2$ distortions to the data simultaneously. Since both cause 
$\mprim = 1$ periodicities in the line-of-sight velocities, however, 
the resulting model
 had too much freedom and produced unphysically large variations in 
all the velocity profiles.

\subsection{Uncertainties}
\label{uncert}
The curvature of the \chisq\ surface at the minimum implies small
formal statistical errors on the best fitting model parameters because
of the large numbers of data values.  The $\chisqm + 1.0/\dof$ contour
on the \chisq\ surface corresponds to variations $\dv < 1\,\kms$ on
the velocity profile points, which we regard as unrealistically small.
We therefore use a bootstrap technique to derive more reasonable
estimates of the scatter in the model parameters about their optimal
values.

 For each model, we generate a bootstrap sample of the data
by adding randomly drawn
residuals $\res\ = \dn\ - \sumk$ from the distribution at
$\chisq=\chisqm$ to the optimal model velocity field.  Since $\{ \res\
\}$ is correlated over a characteristic scale corresponding to $J$
adjacent datapoints, fully random selections do not reproduce the
quasi-coherent residuals we observe in the data.  We therefore select
$P=N/J$ values of \res\ and add them to the model at $P$ random
locations drawn from $\{\rn,\td\}$; residuals at the remaining
$(1-1/J)N$ in $\{\rn,\td\}$ are fixed to the value of the nearest
randomly drawn \res.  We find that $J=4$ produces bootstrap residual
maps with features on scales similar to those in $\{ \res\ \}$ for the
models of \gal\ in Fig.~\ref{models} (see below), but that there is
little change in the derived uncertainties for $2 \leq J \leq 5$.

For both the \rotbi\ and \rotrad\ models, we therefore construct
bootstrap samples of the observed velocity field using $J=4$.  We
repeat the minimization for each sample, substituting the bootstrap
velocities for $\{\dn\}$ in eqs.~\ref{chieq} and \ref{minchieq}.  We
carry out this procedure 1000 times, and adopt the standard deviation
of each parameter about its mean value from all the realizations as
its $1\sigma$ uncertainty in the model of the measured velocities
$\{\dn\}$.

\begin{figure*}
\epsscale{0.9}
\plotone{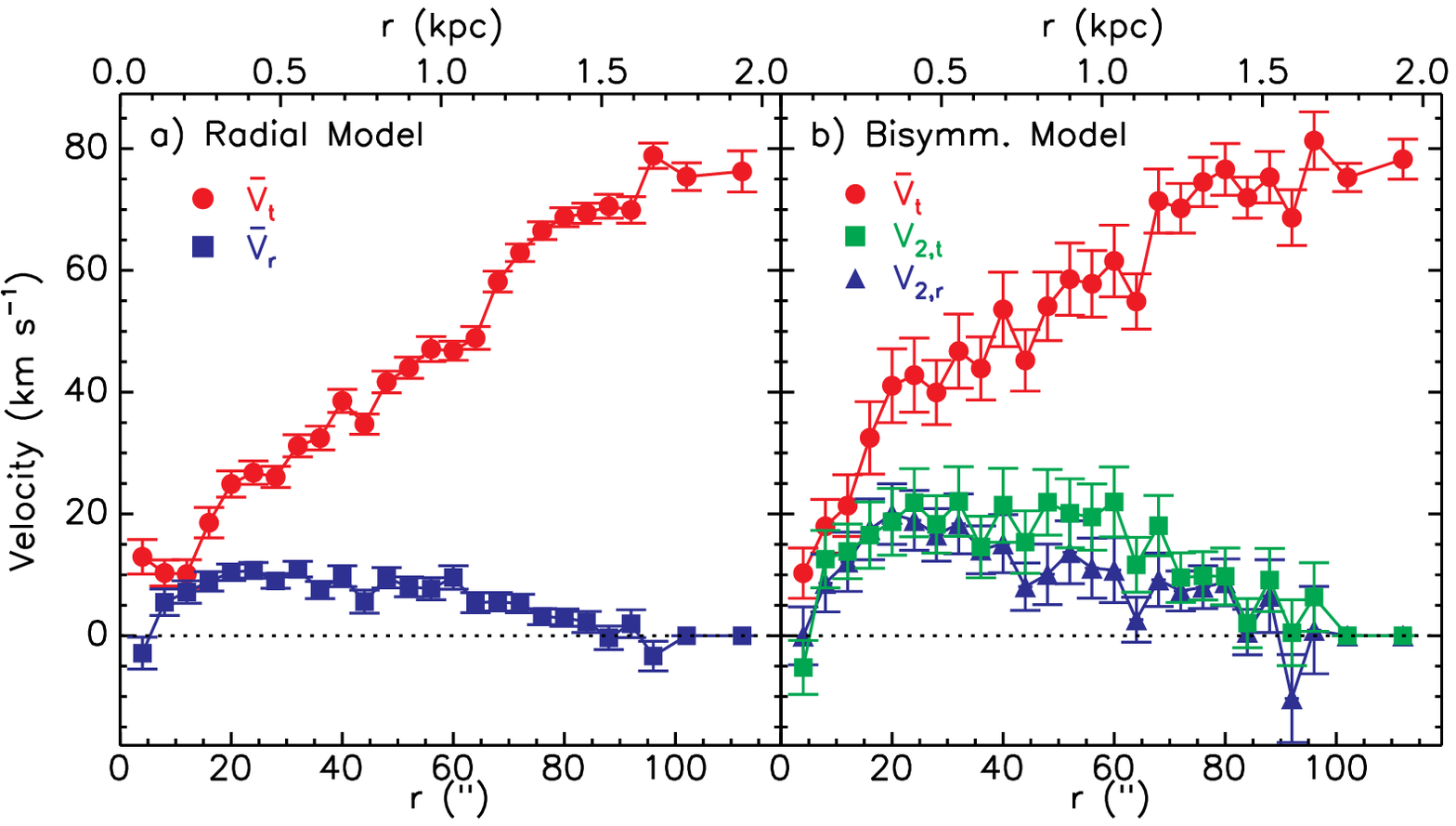}
\caption{Fitted velocity components for \gal.  ({\it a}) The
components of the optimal \rotrad\ model: \Vrotr\ is shown by the red
circles and \Vradr\ is shown by the blue squares (eq.~\ref{radeq}).
({\it b}) Velocity components from the optimal \rotbi\ model: here
\Vrotr\ is shown by the red circles, \Vbitr\ by the green squares, and
\Vbirr\ by the blue triangles (eq.~\ref{bieq}).  \Ignore{See
Fig.~\ref{bicomp} for a plot of $\Vbitr\ - \Vbirr$.  The solid black
line shows a power-law fit to \Vrotr, which corresponds to a density
profile $\rhot \propto r^{-0.39}$ under spherical symmetry. For the
\rotbi\ model (\ref{rcs}{\it b}).  The solid black line shows a
power-law fit to \Vrotr, which corresponds to $\rhot \propto
r^{-1.04}$ under spherical symmetry.}}
\label{rcs}
\end{figure*}

\subsection{Results}
\label{results}
Our final models of the \citetalias{s03} \ha\ and CO velocity fields
for \gal\ are shown in Fig.~\ref{models}.  The minimization results
are given in Table~\ref{fits}, and the corresponding velocity profiles
are shown in Fig.~\ref{rcs}.

The observed velocity field from \citetalias{s03} is reproduced in
Fig.~\ref{models}{\it a}, the best fitting \rotrad\ and \rotbi\ models
are in Figs.~\ref{models}{\it b} and \ref{models}{\it c}, and the
residuals $\{\res\}$ are in Figs.~\ref{models}{\it d} and
\ref{models}{\it e}.  Both models reproduce the gross features of the
observed velocity field, although the \rotbi\ model exhibits a somewhat
larger isovelocity contour ``twist'' along the kinematic minor axis
(oriented vertically in Fig.~\ref{models}) than the \rotrad\
model. The residual patterns in Figs.~\ref{models}{\it d} and
\ref{models}{\it e} are very similar: \res\ is correlated on scales of
$15-20\arcsec$ ($250-350\,$pc) in the maps, which may reflect
large-scale turbulence.  The mean values \averes\ in Table~\ref{fits}
are slightly lower for the \rotbi\ model than for the \rotrad\ one, as
is also suggested by the colors in Figs.~\ref{models}{\it d} and
\ref{models}{\it e}.

The values of \cen\ and \Vsys\ in the two models (Table~\ref{fits})
are identical within their uncertainties, while the \rotrad\ model
favors a larger \eps\ and \pdsky\ than the \rotbi\ model at
the 2$\sigma$ level. Both sets of kinematic parameters $\left(
\xc,\,\yc,\,\eps,\,\pdsky \right)$ are consistent with the photometric
values derived by \citetalias{s03}, corroborating their conclusion
that there is little evidence for an offset between them (see also \S\ref{n2976}). The values
of \chisqm\ indicate that both models adequately describe $\{\dn\}$,
with $\chi^2 \sim 1$ per degree of freedom for the adopted
\eISM.\footnote{The optimal model parameters remain unchanged with
choices of \eISM\ in the range $3\,\kms \lesssim \eISM \lesssim
7\,\kms$, but \chisqm\ of the corresponding fits varies from $2.7
\lesssim \chisqm \lesssim 0.9$.} Even though the \rotbi\ model has
fewer \dof\ than the \rotrad\ model, the difference between their
\chisqm\ is formally significant at the $12\sigma$ level. As with the
unrealistically small model uncertainties implied by the $\chisqm +
1.0/\dof$ contour on the \chisq\ surface, however, a literal
interpretation this difference in goodness-of-fit is unwise. We thus
conclude conservatively that the smaller \chisqm\ and lower \averes\
of the \rotbi\ over the \rotrad\ model imply only a marginally
superior statistical fit to the data.

\begin{figure}
\epsscale{1.1} 
\plotone{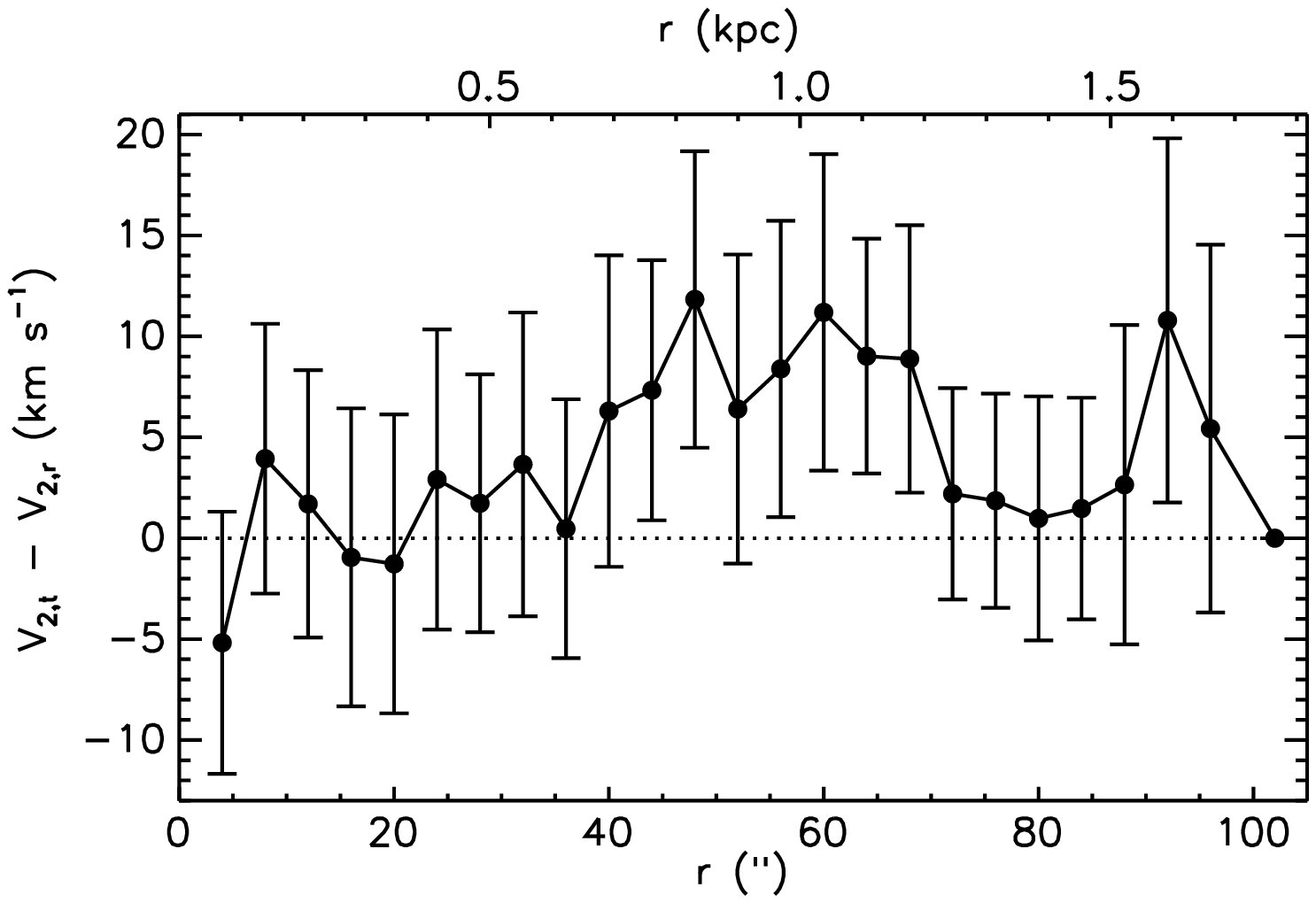}
\caption{Difference $\Vbitr\ - \Vbirr$ in the optimal \rotbi\
model. The components are plotted separately in Fig.~\ref{rcs}{\it
b}.}
\label{bicomp}
\end{figure}

\begin{figure}
\epsscale{1.1}
\plotone{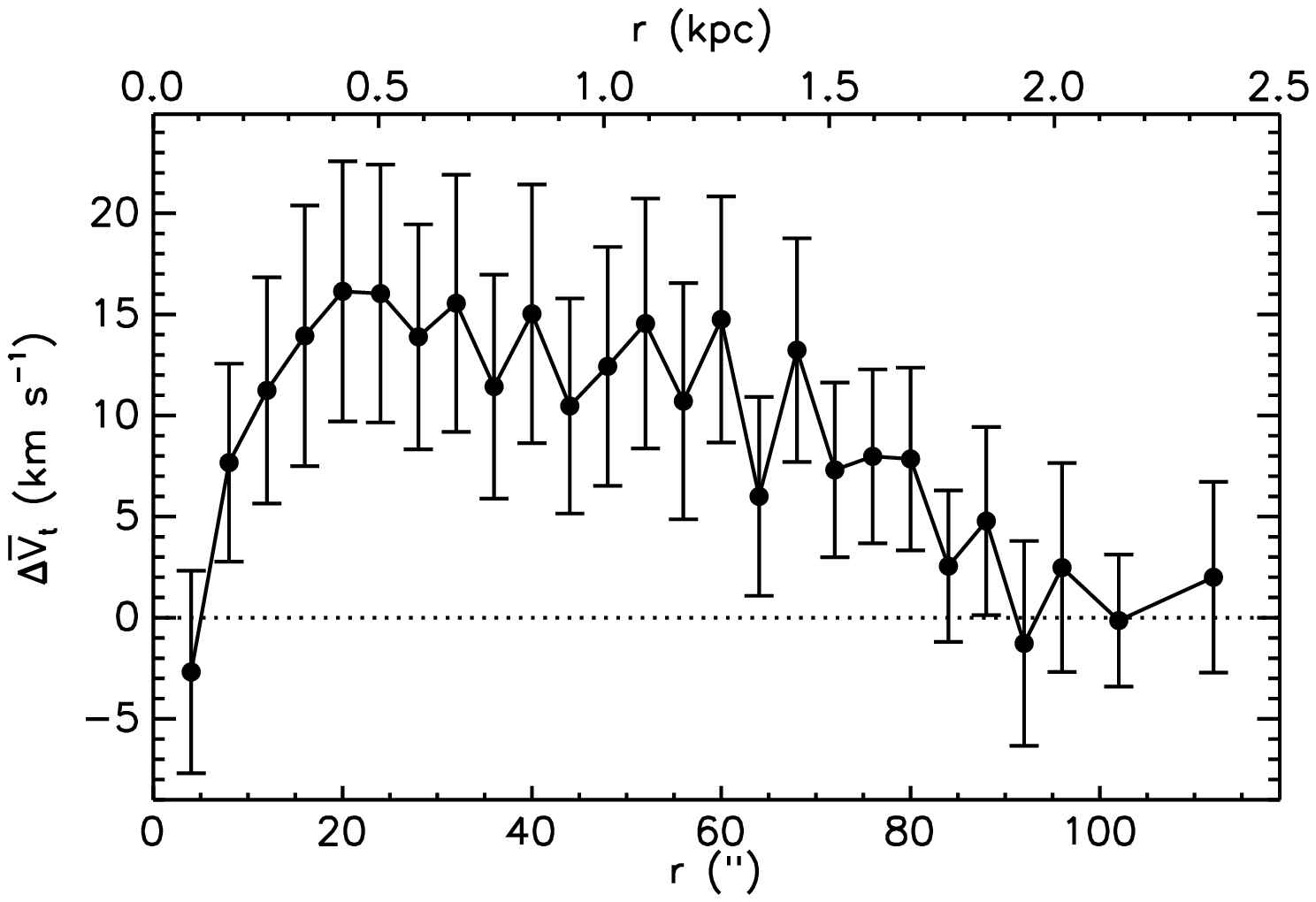}
\caption{Difference \dVrotr\ between the optimal \Vrotr\ from the
\rotbi\ model and the optimal \Vrotr\ from the \rotrad\ model. The
components are plotted separately in Fig.~\ref{rcs}.}
\label{rccomp}
\end{figure}

The best fitting velocity field components from our \rotrad\ model are
shown in Fig.~\ref{rcs}{\it a}.  Despite significant differences
between our minimization technique and that of \citetalias{s03}, our
measurements of \Vrotr\ and \Vradr\ agree well with their
results (the large \Vrotr\ for $r \lesssim 10\arcsec$ in our
\rotrad\ model was also found by \citetalias{s03} in their {\it ringfit}
velocity field decompositions; see their fig.~7a). We find that \Vradr\ is $\sim 7\,\kms$ smaller at $r<
30\arcsec$ than the radial velocity amplitudes presented by
\citetalias{s03}.  This $\sim2\sigma$ discrepancy results from our
inclusion of a \eISM\ term in $\{\sn\}$ (eqs.~\ref{chieq} --
\ref{minchieq}), and disappears if we set $\eISM = 0$ in our model.
Thus our analysis confirms the non-circular motions in \gal\ found by
these authors.

The \rotbi\ model favors a strongly non-axisymmetric flow about an
axis inclined $17^\circ$ to the projected major axis in the disk plane.  The radial
variations and uncertainties of all three fitted velocity components
are shown in Fig.~\ref{rcs}{\it b}.  The estimated uncertainties on
the $\{\vk\}$ are larger than those in the \rotrad\ model
(Fig.~\ref{rcs}{\it a}), consistent with the larger scatter in the
profile values from ring to ring.  This is likely due to the extra
velocity profile relative to the \rotrad\ model, which gives the
\rotbi\ model increased flexibility to fit small-scale features.  As
in the \rotrad\ model, we find significant non-circular motions, this
time in the form of a bisymmetric flow pattern in the disk plane.  The
overall shape of the non-circular contributions \Vbitr\ and \Vbirr\
resembles that of \Vradr\ in the \rotrad\ model: this is reasonable
because both models must fit the $\mprim=1$ variations of the velocity
field.  The difference $\Vbitr\ - \Vbirr$ between the bisymmetric
components in Fig.~\ref{rcs} is plotted in Fig.~\ref{bicomp}. There is
 marginal evidence that $\Vbitr > \Vbirr$ for $45\arcsec
\lesssim r \lesssim 65\arcsec$, but elsewhere the two components have
very similar amplitudes.  Linear theory applied to a weak, stationary
bar-like distortion produces $\Vbitr = \Vbirr$ for a solid-body
rotation velocity profile and $\Vbitr = 1.5\Vbirr$ for a flat one
\citep{sw93}.  Although linear theory cannot be trusted for strong 
perturbations, it is somewhat reassuring that it predicts similar 
\Vbitr\ and \Vbirr\ for a rising \Vrotr.

The most significant difference between the optimal \rotrad\ and
\rotbi\ models is in the shape of \Vrotr.  Beyond the region affected
by non-circular motions, $r \gtrsim 80\arcsec$, \Vrotr\ is identical
in the two models, as it must be, but large differences arise where
non-circular motions are large.  Fig.~\ref{rccomp} shows the
difference between \Vrotr\ from the \rotbi\ model and that from the
\rotrad\ model: the former profile rises more steeply than the latter,
and its amplitude is larger by $\sim 15\,\kms$ for $15\arcsec \lesssim
r \lesssim 50\arcsec$.  We discuss the reason for these differences
in the next section.

\begin{deluxetable*}{lccccccccc}
\tablewidth{0pt}
\tablecaption{Minimization Results \label{fits}}
\tablehead{ \colhead{Model} & \colhead{\eps} & \colhead{\pdsky} & \colhead{\xc} & \colhead{\yc} & \colhead{\Vsys} & \colhead{\pbsky}  & \colhead{\chisqm} & \colhead{\dof} & \colhead{\averes}  \\
       & &\colhead{(\degr)}& \colhead{(\arcsec)}& \colhead{(\arcsec)}& \colhead{(\kms)} & \colhead{(\degr)} &    &   &   \colhead{(\kms)} \\
 \colhead{(1)} & \colhead{(2)} & \colhead{(3)} & \colhead{(4)} & \colhead{(5)} & \colhead{(6)} & \colhead{(7)} & \colhead{(8)} & \colhead{(9)} & \colhead{(10)} }
 \startdata
 \rotrad\ & $0.568 \pm 0.007$  & $-36.0 \pm 0.6$ & $-1.7 \pm 0.3$ & $0.8 \pm 0.4$ & $0.3 \pm 0.4$ & \nodata   & 1.35  & 1034 & 3.4  \\
 \rotbi  & $0.556 \pm 0.007$  & $-37.6 \pm 0.6$ & $-1.9 \pm 0.3$ & $1.2 \pm 0.3$ & $0.7 \pm 0.4$ & $-45 \pm 4$ & 1.20  &  1009 & 3.1  \\
 \enddata
 \tablecomments{Col. (1): Model. Col. (2): Disk ellipticity. Col. (3): Disk PA, measured North $\rightarrow$ East to the receding side of the disk. Col. (4): Right ascension of disk center, relative to the photometric center $09^\mathrm{h}\,47^\mathrm{m}\,15\fs\,3$ \citepalias{s03}. Col. (5): Declination of disk center, relative to photometric center $67\degr\,55\arcmin\,00\farcs\,4$ \citepalias{s03}. Col. (6): Disk systemic velocity, heliocentric optical definition. Col. (7): Bisymmetric distortion PA, measured North $\rightarrow$ East. Col. (8): Minimum value of \chisq\ (eq.~\ref{chieq}) obtained. Col. (9): Number of degrees of freedom in the minimization. Col. (10): Amplitude of the average (data - model) residual.}
 \end{deluxetable*}

\section{Discussion}
\label{discuss}

\subsection{Mean streaming speed}
\label{streaming}
The large differences in the fitted \Vrotr\ for the bisymmetric and 
radial models over the
inner part of \gal\ (Fig.~\ref{rccomp}) demand explanation.  Fig.~\ref{comps} shows
sky-plane projections of the angular variations of the separate fitted
velocity components at $r=20\arcsec$ for both models.  

Fig.~\ref{comps}{\it a} shows the projected \Vradr\ of the \rotrad\
model (dash-dotted line), which shifts the peak in the
projected model velocity (solid line) away from the kinematic
major axis ($\td = 0$) and reproduces the iso-velocity ``twist'' in the
observed velocity field (Fig.~\ref{models}). Since the projected
\Vrad\ must be zero at $\td=0$ in this model (see eq.~\ref{radeq}),
the projected \Vrotr\ (dashed line) must equal \Vmod\ along the
kinematic major axis.

Fig.~\ref{comps}{\it b} shows the corresponding case for the \rotbi\
model where the non-axisymmetric terms allow \Vmod\ to differ from
\Vrot\ along the major axis (eq.~\ref{bieq}).  The abscissae are
marked as \td\ along the bottom and as \tb\ along the top, which
differ only slightly because the best fitting major axis of the
bisymmetric distortion in \gal\ projects to a PA similar to the
kinematic major axis (i.e. $\pbsky \sim \pdsky$; 
Table~\ref{fits}).  The significant negative contribution of the
projected \Vbitr\ (dash-dotted line) to \Vmod\ (solid line)
at $\td = 0$ offsets the positive contribution from \Vrotr\ (dashed
 line).  The greater amplitude of \Vrotr\ in the \rotbi\ model of
\gal\ is therefore due to the large non-circular motions in the inner
parts that happen to be negative near the kinematic major axis because
the $m=2$ distortion is oriented close to this axis.

Notice also that the \Vbit\ and \Vbir\ components in Fig.~\ref{comps} show,
 as they must
(\S\ref{theory}), both $\mprim = 1$ and $\mprim =3$ periodicities, and
that both are of similar amplitude (see also Fig.~\ref{bicomp}).  
Yet their relative phases ensure
that the net effect of the $\mprim=3$ terms on \Vmod\ cancels almost
exactly.  A larger $\mprim = 3$ signal could arise if the \Vbit\ and
\Vbir\ terms have different amplitudes, but their relative phases
always ensure at least partial cancellation regardless of the
orientation of the projected bar.  Thus one should not conclude that a
very weak $\mprim=3$ signal in the velocity map implies no significant
bisymmetric distortion.

\subsection{Centrifugal balance?}
\label{balance}
It is clear from Table~\ref{fits} and Fig.~\ref{models} that both the
\rotbi\ and \rotrad\ models are adequate parameterizations of the
observed geometry and kinematics of \gal.  But does either model
provide insight into its physical structure?

The mean orbital speed, \Vrotr, in the \rotrad\ model can balance the
central attraction of the system only if either the non-circular
motions are small, or \Vradr\ actually implies a real radial flow that
somehow does not affect orbital balance.  The first possibility is not
true, as we have confirmed (Figs.~\ref{rcs} and \ref{comps}) the large
non-circular motions found for $r \lesssim 500\,$pc in \gal\ by
\citetalias{s03}. If \Vradr\ is attributed to radial flows that do not
affect orbital balance, then all of the detected gas in this quiescent
system would be displaced on kpc scales in $1-3\,$Gyr; we agree with
\cite{s05} that this explanation is not viable.  We thus conclude that
although the optimal \rotrad\ model is a reasonable statistical fit to
the data and provides strong evidence for non-circular motions, the
fitted \Vrotr\ cannot be used to determine the mass distribution
within \gal.

If the non-circular motions in \gal\ are dominated by an $m=2$
perturbation to its potential, then the velocity profiles of the
optimal \rotbi\ model should better reflect the galaxy's structure
than those of the \rotrad\ model.  While the fitted \Vrotr\ rises more
steeply in the \rotbi\ model, it is merely the average azimuthal speed
around a circle, not a precise indicator of circular orbital
balance.  It should be stressed that circles in the disk plane
approximate streamlines only when non-circular motions are small.  In
a bar-like potential, the gas on the bar major axis will be moving
more slowly than average, since it is about to plunge in towards the
center, whereas gas at the same galactocentric radius on the bar minor
axis will be moving faster than average, since it has arrived there
from a larger radius.  Under these circumstances, it is not possible
to assert that the azimuthal average, \Vrot, is exactly equal to the
circular orbit speed in an equivalent azimuthally averaged mass
distribution.  The only reliable way to extract the azimuthally
averaged central attraction in this case is to find the
non-axisymmetric model that yields a fluid dynamical flow pattern to
match that observed, and to average afterwards.

Despite these cautionary statements, we suspect that the \Vrot\ curve from the \rotbi\ model provides a better estimate, than does that from the \rotrad\ model, of the azimuthally averaged central attraction in \gal.

% Despite these cautionary statements, we suspect that \Vrotr\ for the
%%that results from centrifugal 
%balance with the azimuthally averaged mass distribution than \Vrotr\ 
%for the \rotrad\ model. 

\subsection{Evidence for a bar}
\label{bar}

As discussed in \S\S\ref{intro} \& \ref{technique}, the elliptical
streams of fixed direction and phase in the \rotbi\ model could be
driven by either a triaxial halo or by a bar in the mass distribution.
In either case, the distortion is significant only at $r \la 80\arcsec$ 
($1.4\,$kpc; Fig.~\ref{rcs}{\it b}) in \gal, beyond 
which the flow appears to be near circular.

The aspherical halo interpretation therefore requires the halo that
hosts \gal\ to have an asphericity that increases for decreasing $r$.
Such an idea was proposed by \citet{hayashi06new}, although other work
\citep{dub94,gnedin04,kazant04,beren06,gustaf06} has indicated a tendency for disk
assembly to circularize the potential.

We therefore favor the interpretation that \gal\ hosts a bar.
\citet{kmd07} have examined the {\it Two Micron All Sky Survey\/}
\citep[2MASS;][]{skrutskie06} $J$, $H$ and $K_s$ images of \gal\ to
search for a bar.  Their fits to this photometry reveal a radial variation in
ellipticity of amplitude $\Delta\epsilon > 0.1$ \citep[see
also][]{s05}, and their visual inspection of the images reveals a ``candidate" bar
with $\pa_{\rm bar}=-43 \degr$ and semi-major axis $a = 72 \pm
5\arcsec$ (see their table~2).  Their estimated $\pa_{\rm bar}$ is
fully consistent with our kinematic estimate \pbsky\ (Table~\ref{fits}), 
and $a$ compares well with the range of $r$ where
\Vbitr\ and \Vbirr\ are non-zero (Fig.~\ref{rcs}{\it b}). Furthermore,
$\pa_{\rm bar}$ and \pbsky\ are roughly coincident with the apparent
major axis of the CO distribution
%\footnote{We have verified that
%\rotbi\ model fits to the \ha\ kinematics alone imply an $m=2$
%distortion to the potential similar to that presented in
%Table~\ref{fits}.} 
in \gal\ (see fig.~4 of \citetalias{s03}), which
suggests that the molecular gas density is larger along this PA than
elsewhere in the disk.  Thus the 2MASS photometry and CO morphology
provide strong supporting evidence that \gal\ contains a bar with the
properties implied by our \rotbi\ model.

\subsection{Mass components}
\label{mass}
Our fits have revealed strong non-circular motions in \gal\ that appear to
result from forcing by a bar.  While \Vrotr\ in the \rotbi\ model 
better reflects the azimuthally averaged mass distribution than its 
counterpart in the \rotrad\ model, precise statements about the mass budget 
in \gal\ are hampered by our lack of a reliable
estimate of the circular orbital speed curve (see \S\ref{balance}).

The amplitude of the non-circular motions in \gal\ implies a
relatively large bar mass, which in turn suggests that the disk itself
contributes significantly to the central attaction.  It is therefore
likely that the baryons in \gal\ dominate its kinematics well beyond
the $r \sim 500\;$pc suggested by the fits of \citetalias{s03}.
Indeed, the steeper rise of \Vrotr\ in the \rotbi\ model relative to
that deduced by \citetalias{s03} would allow a larger disk
mass-to-light ratio (${\cal M}/L$) to be tolerated by the kinematics.
This conclusion eases the tension between their dynamical upper bound
on the stellar ${\cal M}/L$ and that expected from stellar population
synthesis for the observed broadband colors \citepalias[see \S3.1.1
of][]{s03}.

We defer the detailed mass modeling of \gal\ required for quantitative
estimates of its mass budget to a future paper.  Such a study would be
assisted by additional kinematic data from extant \hi\ aperture
synthesis observations, as well as by decompositions of publicly available infrared photometry from the {\it Spitzer Infrared Nearby Galaxies Survey} \citep{sings}.

\subsection{Other galaxies}
\label{othergals}
We suggest that our approach could be useful for characterizing the
non-circular motions detected in other galaxies, particularly in low-mass
systems where the reported non-circular motions are large 
\citep{swaters03,s05,gentile06}.  It is more direct than
interpretations of velocity field Fourier coefficients in the weak
perturbation limit, yields physically meaningful kinematic components
for systems with bar-like or oval distortions to the potential, and
its application is much simpler than that of a full fluid-dynamical
model.

%We have shown that the velocity field of \gal, when fitted by our
%bisymmetric model, reveals a steeper inner rise in \Vrotr\ than in
%previous analyses by other methods.  The reason for this difference
%(Fig.~\ref{comps}) is that the \Vbit\ terms happen to partly cancel the
%\Vrot\ terms, because they have opposite signs on the projected major
%axis when the bar is oriented close to this direction.  It should be
%clear, however, that the \Vbit\ terms will have the opposite effect if
%the bar is more nearly aligned with the projected minor axis.  Thus
%even if the non-circular motions detected in other systems result from
%bars in the potential, it is unlikely that the our \rotbi\ model will
%always cause \Vrotr\ to rise more steeply than found previously.  In
%any event, it should be clear that when large non-circular flows are
%present, the mean orbital speed derived from models that use epicycle
%theory can yield a very misleading estimate of the interior mass
%needed for centrifugal balance.

We have shown that the velocity field of \gal, when fitted by our
bisymmetric model, reveals a steeper inner rise in \Vrotr\ than in
previous analyses by other methods. Similar findings have been 
reported by Hayashi \& Navarro (2006) and Valenzuela et al. (2007) 
for other systems. In \gal, the reason for this difference
(Fig.~\ref{comps}) is that the \Vbit\ terms happen to partly cancel the
\Vrot\ terms, because they have opposite signs on the projected major
axis when the bar is oriented close to this direction. It should be
clear, however, that the \Vbit\ terms will have the opposite effect if
the bar is more nearly aligned with the projected minor axis.  Thus
even if the non-circular motions detected in other systems result from
bars in the potential, it is unlikely that our \rotbi\ model will
always cause \Vrotr\ to rise more steeply than found previously.  In
any event, it should be clear that when large non-circular flows are
present, the mean orbital speed derived from models that use epicycle
theory can yield a very misleading estimate of the interior mass
needed for centrifugal balance.

\section{Conclusions}
\label{conclusions}

We have presented a new method for fitting 2-D velocity maps of spiral
galaxies that are characterized by non-circular motions.  We suppose
the potential to contain a bar-like or oval distortion that drives the
gas in the disk plane on an elliptical flow pattern of fixed
orientation, such as could arise from a triaxial halo or a bar in the
mass distribution of the baryons.  Our model has important advantages
over previous approaches since it is not restricted to small
non-circular motions, as is required when epicycle theory is employed,
and we do note invoke large radial flows that have no clear physical
origin or interpretation.

Our \rotbi\ flow model can be fitted to data by a generalization of
the technique developed by \citet{bs03}.  The fit extracts multiple
non-parametric velocity profiles from an observed velocity field,
and we employ a bootstrap method to estimate uncertainties.

\begin{figure*}
\epsscale{0.9}
%\plotone{f6.eps}
\plotone{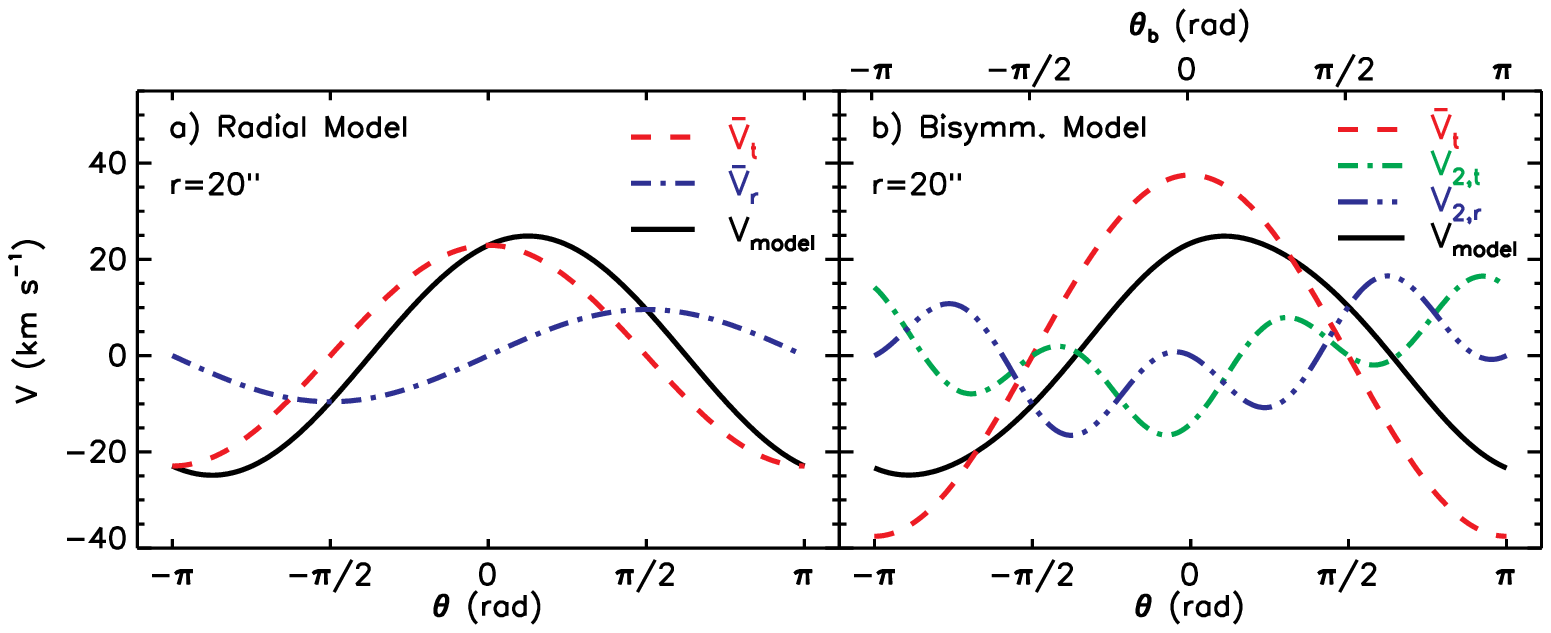}
\caption{Projected contributions from different kinematic components
at $r=20\arcsec$ ($345\,$pc) in the optimal ({\it
a}) \rotrad\  and ({\it b}) \rotbi\  models. In \ref{comps}{\it a},
the dashed line shows the angular dependence of the projected rotational
velocity term in the \rotrad\ model relative to the kinematic major axis
 ($2^\mathrm{nd}$
on the right-hand side (RHS) of eq.~\ref{radeq}), and the
dash-dotted line shows that of the radial velocity term
($3^\mathrm{rd}$ on the RHS of eq.~\ref{radeq}). In \ref{comps}{\it
b}, the angular dependence of the components in the \rotbi\ model are
plotted relative to \pdsky\ along the bottom horizontal axis and \pbsky\
along the top horizontal axis. The dashed line shows the
rotational velocity term ($2^\mathrm{nd}$ on the RHS of
eq.~\ref{bieq}), and the dash-dotted and  dash-dot-dotted
lines show the tangential and radial bisymmetric terms, respectively
($3^\mathrm{rd}$ and $4^\mathrm{th}$ on the RHS of
eq.~\ref{bieq}). The solid lines in both panels shows the net
projected model velocity relative to \Vsys.  }
\label{comps}
\end{figure*}

As an example, we have applied our technique to the \ha\ and CO
kinematics of \gal\ presented by \citetalias{s03}.  We show that the
\rotbi\ model fits the data at least as well as the {\it ringfit}
procedure implemented by these authors that invokes large radial
velocities, which we are also able to reproduce by our methods.  Both
the \rotbi\ and \rotrad\ models reveal large non-circular motions in
\gal, but the derived mean orbital speed profiles \Vrotr\ differ
markedly between the two cases.  We explain the reason for this large
difference in \S\ref{discuss}.

When disks are observed in projection, kinematic distortions with intrinsic sectoral harmonic $m$ cause azimuthal variations of orders $\mprim = m \pm 1$ in line-of-sight velocity maps.
Our analysis of \gal\ clearly demonstrates that $\mprim = 1$
distortions to its velocity field can be fitted by a bisymmetric
distortion to the potential, which we regard as more physically
reasonable than radial flows.  We show in Fig.~\ref{comps} that
$\mprim = 3$ distortions should be small in the \rotbi\ model; this is
because the $\mprim = 3$ variations in the radial and tangential
components project out of phase.  They will cancel exactly only when
of equal amplitude, which should be approximately true in the rising
part of \Vrotr.

We suggest that \gal\ hosts a strong bar oriented at $\sim
17^\circ$ to the projected major axis.  Our interpretation is
supported by its CO morphology \citepalias{s03} and more strongly by
the results of \citet{kmd07}, who analyzed the 2MASS photometry of
\gal\ and found a bar whose size and orientation are similar to those
required by our \rotbi\ model.

We find that the mean orbital speed in \gal\ rises more
steeply than indicated by previous studies (\citetalias{s03};
\citealt{s05}).  While \Vrotr\ in our \rotbi\ model 
is not an exact measure of the
circular orbital speed in the equivalent axially symmetrized galaxy, we regard
it as a better approximation to this quantity.  Since the strongly
non-circular flow pattern implies a massive bar, which in turn
suggests a massive disk, we expect a larger baryonic mass than was
estimated by \citetalias{s03}.  It is likely, therefore, that most of
the increased central attraction required by our more steeply rising
\Vrotr\ will not reflect a corresponding increase in the density of the
inner dark matter halo, but will rather ease the tension between
maximum disk fits to its kinematics and ${\cal M}/L$ predictions from
broadband photometry \citepalias{s03}.  Indeed, since non-circular
motions are detected throughout the region $r \lesssim 80\arcsec$
($1.4\,$kpc), it seems likely that the luminous matter in \gal\ is an
important contributor to the central attraction at least as far out as
this radius.  Detailed mass models of this system are forthcoming.

Application of our method to other galaxies will not always result in
a steeper inner rise in the mean orbital speed.  We find this
behavior in \gal\ only because the bar is oriented near to the
projected major axis.  Neglect of non-cirular motions, or application
of a radial flow model, when the bar is oriented close to the
projected minor axis will lead to an erroneously steep rise in the
apparent inferred mean orbital speed, 
which will rise less steeply when our model is applied.

\acknowledgments 
We thank Josh Simon for providing the data for
\gal, and Alberto Bolatto for help in interpreting the 
measurement uncertainties. We also thank Alberto Bolatto and Josh Simon for helpful comments on the manuscript. KS is a Jansky
Fellow of the National Radio Astronomy Observatory.  JAS is partially
supported by grants from the NSF (AST-0507323) and from NASA
(NNG05GC29G).

\clearpage

\appendix

\section{Kinematic Model Weights}
\label{appen}

We use the same notation and parameter definitions as in
\S\ref{technique} and Fig.~\ref{setup}.  Let \ptn\ be the location, in
the sky plane, of the $n$th measured velocity \dn, where
$\vec{\hat{x}}$ points West and $\vec{\hat{y}}$ points North.  Let
\pdsky\ be the \pa\ of the projected major axis. The coordinates \pte\
centered on \cen\ and aligned with the projection are:
\begin{eqnarray}
\nonumber
\xe&=&-(\xn-\xc)\sin{\pdsky} + (\yn-\yc)\cos{\pdsky}\\
\ye&=&-(\xn-\xc)\cos{\pdsky} - (\yn-\yc)\sin{\pdsky}.
\label{rotframe}
\end{eqnarray}
The radius \rn\ of the circle in the disk plane that passes through
the projected position of \ptn\ is given by
\begin{equation}
\rn^2 = \xe^2 + \left( \frac{\ye}{1-\eps} \right)^2\;,
\label{sma}
\end{equation}
where \eps\ is the ellipticity of the circle caused by projection;
i.e.\ $\eps = 1 - \cos i$ for the thin gas layer considered here, where $i$ 
is the galaxy inclination.
Thus the \pa, \td, of the measurement at \ptn\ relative to the disk
major axis (see Fig.~\ref{setup}) satisfies
\begin{equation}
\cos{\td}=\frac{\xe}{\rn}\;,
\label{costheta}
\end{equation}
\begin{equation}
\sin{\td}=\frac{\ye}{(1-\eps)\rn}\;.
\label{sintheta}
\end{equation}

We tabulate each of the non-parametric velocity profiles at a set
of radii in the disk plane that project to ellipses on the sky with
semi-major axes $\{ {a_{k'm}} \}$.  The elements of \vkb\ include all
$K=\sum_M K'_M$ values from the $M$ velocity profiles combined.  In
principle, we could employ different numbers of rings
$K'_M$, with differing choices for the $\{ a_{k'm} \}$ for each velocity
profile, but here we evaluate all
profiles at the same $K'$ locations $\{ a_{k'} \}$.  There are $K'_1$ radii
that define \Vrotr\
in our models (eqs.~\ref{bieq} and \ref{radeq}), but we restrict the number
of rings describing the non-circular components to $K'' < K'$ when the
latter are not well-constrained in the outer disk.  In addition, we
include the systemic velocity \Vsys\ as the $K^\mathrm{th}$ element
of $\{ \vk \}$, with $w_{K,n}=1$ for all \dn.  

We use linear interpolation between the two rings that straddle each
data point.  If $a_k \le \rn < a_{k+1}$, the values of \wkb\ for that
velocity profile and that \dn\ are
\begin{eqnarray}
\nonumber 
w_{k,n} &=& \left( 1 - \frac{\rn - a_k}{\delta a_k} \right) W_n\\ 
\nonumber 
w_{k+1,n} &=& \left(\frac{\rn - a_k}{\delta a_k} \right) W_n\\ 
w_{k',n} &=& 0 \qquad \mathrm{for}\,\,\,\,k^{\prime} \ne k,\,k+1 \,\,\,,
\label{weights}
\end{eqnarray}
where $\delta a_k = a_{k+1} - a_k$ is the ring spacing.  The parameter
$W_n$ in eq.~\ref{weights} is the combination of trigonometric factors
in the dependence of \wkb\ on the projection geometry.  It is clear
from eqs.~\ref{bieq}~and~\ref{radeq} that we require a different
$W_n$ for each \dn\ and for each velocity profile in the \rotbi\ and
\rotrad\ models.

For the \Vrot\ component in both models, we have
\begin{equation}
W_n = \sini \cos{\td}\;,
\label{vrotw}
\end{equation}
where $\cos{\td}$ is given in eq.~\ref{costheta}; for the \Vbit\ component 
in the \rotbi\ model
\begin{equation}
W_n = \sini \cos(2\tb) \cos{\td}\;,
\label{bitw}
\end{equation}
and for the \Vbir\ component
\begin{equation}
W_n = \sini \sin(2\tb) \sin{\td}\;.
\label{birw}
\end{equation}
\Ignore{For a given value for the \pa\ \pbp\ of the bisymmetric
distortion relative to \pdsky, the factor $\cos(2\tb)$ and
$\sin(2\tb)$ can be computed from eqs.~(\ref{costheta}) --
(\ref{sintheta}) via eq.~(\ref{barmajeq}) and standard trigonometric
angle addition formulas. }Finally, $W_n$ for the \Vrad\ component in
the \rotrad\ model is
\begin{equation}
W_n = \sini \sin{\td}\;.
\label{radw}
\end{equation}

\clearpage

\end{document}